\documentclass[a4paper]{jpconf}
\usepackage{graphicx}
\usepackage{lineno}

\begin{document}

\title{Muon Physics in ALICE: The MFT Upgrade Project}

\author{Antonio Uras for the ALICE MFT Working Group}

\address{IPNL, Universit\'e Claude Bernard Lyon-I and CNRS-IN2P3, Villeurbanne, France}

\ead{antonio.uras@cern.ch}


\begin{abstract}
The ALICE experiment is dedicated to the study of the quark gluon plasma in heavy-ion collisions at the CERN LHC. The Muon Forward Tracker (MFT) is under consideration by the ALICE experiment to be part of its program of detectors upgrade to be installed during the LHC Long Shutdown~2 (LS2) planned for 2018. Designed as a silicon pixel detector added in the Muon Spectrometer acceptance ($-4.0 < \eta < -2.5$) upstream of the hadron absorber, the MFT will allow a drastic improvement of the measurements that are presently done with the Muon Spectrometer and, in addition, will give access to new measurements that are not possible with the present Muon Spectrometer setup. Motivations and preliminary results are discussed here, concerning the measurement of prompt and displaced charmonia, open heavy flavors, and low mass dimuons in central Pb--Pb collisions at $\sqrt{s_{NN}} = 5.5$~TeV.
\end{abstract}

\section{Introduction}

Lepton-pair measurements always played a key-role in high energy nuclear physics: leptons arrive to the detectors almost unaffected, being not sensible to the strong color field dominating inside the QCD matter, thus providing an ideal tool to probe the whole evolution of nuclear collisions. The \mbox{ALICE} experiment at the CERN LHC measures single lepton and lepton pair production both at central rapidity, in the electron channel, and at forward rapidity, in the muon channel. Identification and measurement of muons in ALICE are performed in the Muon Arm~\cite{Aamodt:2008zz}, covering the pseudo-rapidity region $-4 < \eta < -2.5$. Starting from the nominal interaction point (IP) the Muon Arm is composed of the following elements: a hadron absorber allowing a reliable muon identification; a dipole magnet providing a field integral of 3~Tm; a set of five tracking stations, each one composed of two cathode pad chambers with a spacial resolution of about $\sim 100~\mu$m in the bending direction; an 
iron wall, which absorbs the residual secondary hadrons emerging from the front absorber; the muon trigger system, consisting of 4~resistive plate chambers, with a time resolution of about 2~ns. The ALICE experiment has already an intense physics program based on muon measurements, developed along three main directions: study of quarkonia production~\cite{Alice:2012nf, Alice:2012fn}, of open heavy flavor production~\cite{Alice:2012aa, Alice:2012bb}, of low mass dimuons~\cite{Alice:2012cc}. 

\section{Motivations for a Muon Forward Tracker in ALICE}

Despite the remarkable results already obtained, the current ALICE muon physics program suffers from several limitations, basically due to the multiple scattering induced on the muon tracks by the hadron absorber. The details of the vertex region are completely smeared out: in particular, this prevents us to disentangle open charm and open beauty production without making assumptions relying on physics models, and makes it impossible to distinguish prompt and displaced $J/\psi$ production. In addition, only very limited possibilities are given to reject muons coming from semi-muonic decays of pions and kaons, representing an important background both in single muon and dimuons analyses, in particular at low masses and low $p_\mathrm{T}$. Finally, the degradation of the kinematics, imposed by the presence of the hadron absorber, plays a crucial role in determining the mass resolution for the resonances, especially at low masses.
To overcome these limitations, and better exploit the unique kinematic range accessible by the ALICE Muon Arm at the LHC, the Muon Forward Tracker (MFT) was proposed in the context of the ALICE upgrade plans, to take place in the years 2017/2018 during the LHC Long Shutdown~2. The MFT is a silicon pixel detector added in the Muon Spectrometer acceptance ($-4.0 < \eta < -2.5$) upstream of the hadron absorber. The basic idea, motivating the integration of the MFT in the ALICE setup, is the possibility to match the extrapolated muon tracks, coming from the tracking chambers \emph{after} the absorber, with the clusters measured in the MFT planes \emph{before} the absorber; the match between the muon tracks and the MFT clusters being correct, muon tracks gain enough pointing accuracy to permit a reliable measurement of their offset with respect to the primary vertex of the interaction. As realistic simulations show, correct matching rates as large as $60\,\%$ and offset resolutions of $\approx 100~\mu$m are 
expected 
already for muons with $p_\mathrm{T} \approx 1$~GeV/$c$; for $p_\mathrm{T} \approx 3$~GeV/$c$, less than $10\,\%$ of the muons is found to be wrongly matched and the offset resolution stays below $70~\mu$m.

\section{Charmonia}

The interest of measuring charmonia production in heavy-ion collisions at low $p_\mathrm{T}$, mainly comes from the possibility to understand the interplay between suppression/recombination mechanisms in the deconfined medium. From this point of view, the MFT is called to give two main contributions: (i) reducing the uncertainties on the extraction of the $\psi'$ signal; (ii) allowing a reliable disentanglement between prompt and displaced charmonia production. 

\begin{figure}[htbp]
  \begin{center}
  \includegraphics[width=.46\textwidth]{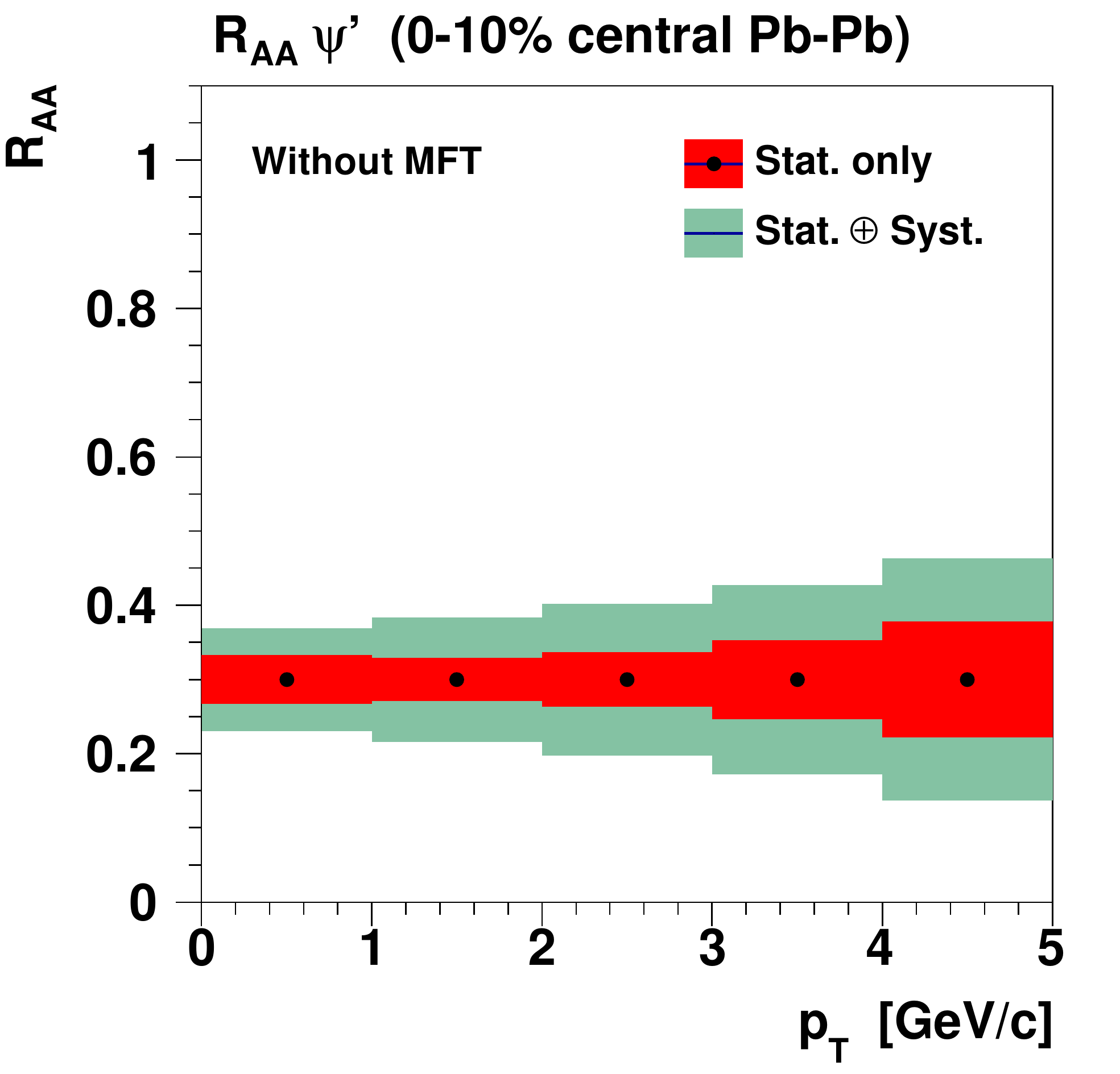} \hspace{0.06\textwidth}
  \includegraphics[width=.46\textwidth]{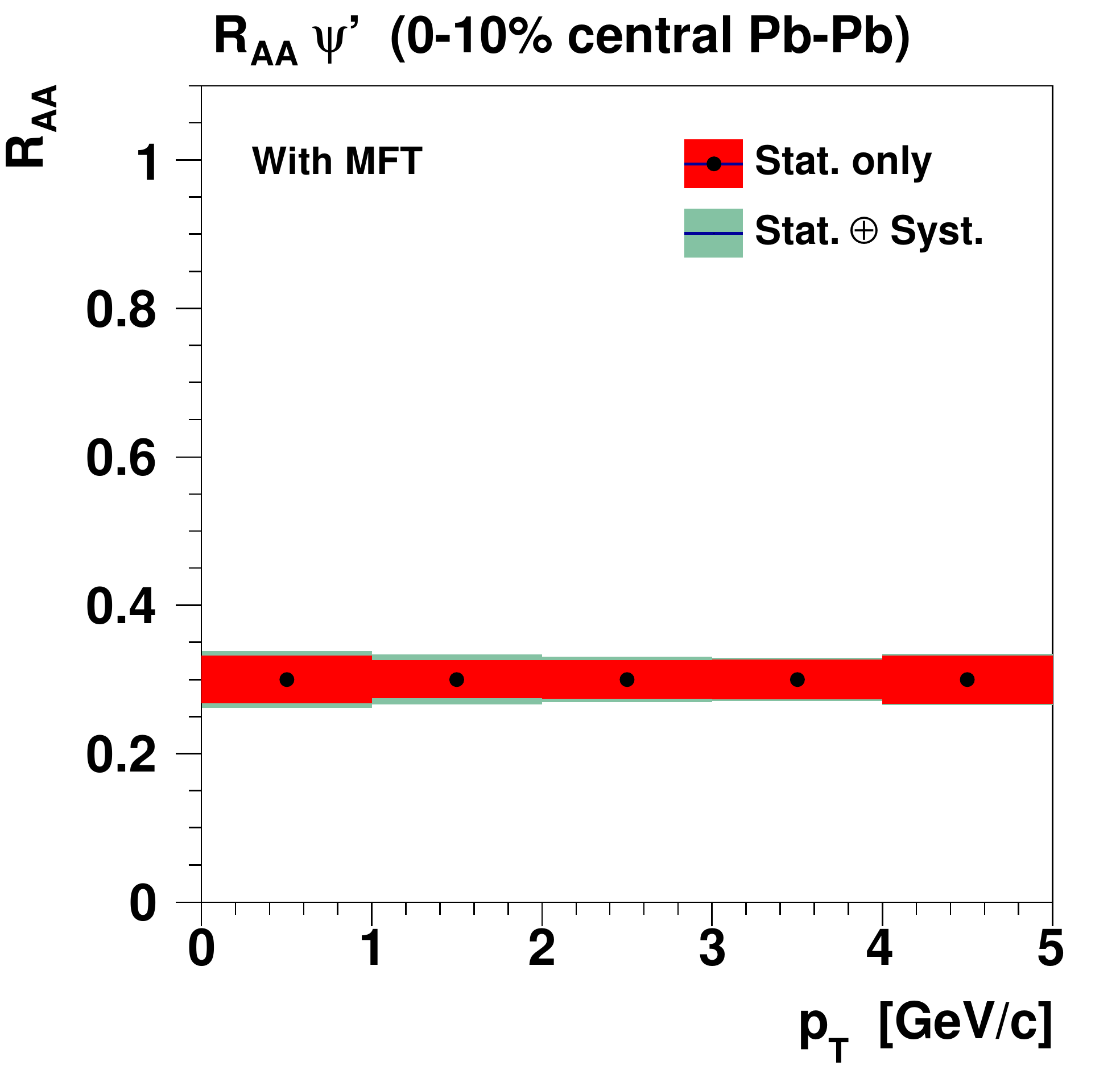} \\
  \end{center}
  \vspace{-0.7cm}
  \caption[\textwidth]{\label{fig:RAA_0.7_0.3} Expected uncertainties on the measure of the nuclear modification factor $R_{AA}$ of the $\psi'$, in the two scenarios without (left panel) and with (right panel) the MFT. The scenario $R_{AA}(\psi') = 0.3$ is considered here. Red, smaller boxes account for statistical uncertainties alone; green, larger ones represent the sum in quadrature of statistical and systematic uncertainties.}
\end{figure}

The main limitation affecting the measurement of the $\psi'$ signal with the current Muon Arm setup comes from the poor signal over background (S/B) ratio, ranging from $0.2\,\%$ to $0.5\,\%$ according to the $p_\mathrm{T}$ range. When comparing these values to the expected systematic uncertainty on the evaluation of the shape and the normalization of the background, estimated from the analysis of the current $J/\psi$ data to be $\sim 0.1\,\%$, one can easily conclude that the only way to improve the uncertainty on the extraction of the $\psi'$ signal is to increase the S/B. The introduction of the MFT in the Muon Arm setup goes exactly in this direction: thanks to its capability to access the details of the vertex region, it makes it possible to operate quality cuts on the dimuon sample, rejecting at the same time: (i) muon pairs in which at least one muon is produced within the hadron absorber, and (ii) muon pairs whose muons are not recognized to share a common origin. Improvement by a factor 3 to~10 is 
found for the S/B and the systematic errors, depending on the $p_\mathrm{T}$; moreover, although the improvement of the S/B is obtained by means of a sensible statistics loss, both the significance and the statistical errors on the isolated signal (after background subtraction) are preserved or even slightly improved. The comparison between the expected statistical and systematic uncertainties, in the two scenarios without and with the MFT, is shown in \figurename~\ref{fig:RAA_0.7_0.3}: we present the results in the form of uncertainties on the nuclear modification factor $R_{AA}$ of the $\psi'$, under the hypothesis $R_{AA} = 0.3$.

Concerning the measurement of the $J/\psi$ production from beauty mesons, it should be remarked that this kind of measurement is simply not accessible with the current Muon Arm setup. The MFT will make it affordable thanks to the possibility to disentangle prompt and displaced $J/\psi$ vertices: the separation between the two components is performed on a statistical basis by means of the analysis of the distributions of the pseudo-proper decay length of the dimuons falling within the $J/\psi$ mass window, defined as $l_{J/\psi} = L_{xy} m_{J/\psi}/p_\mathrm{T}$, where $L_{xy}$ is the most probable transverse decay length of the $b$-hadron, in the laboratory frame. This latter is evaluated as $L_{xy} \approx \frac{\hat{u}^T \cdot \vec{r}}{\hat{u}^T \cdot \hat{u}}$, where $\vec{r}$ represents the transverse displacement vector between the primary vertex and the one of the $\mu^+\mu^-$ pair, and $\hat{u}$ is the unit vector in the direction of the $J/\psi$. Technically, the analysis is performed by decomposing 
the measured pseudo-proper decay length into the three components coming from background dimuons, prompt and displaced $J/\psi$; the normalization of the background component being conveniently fixed by means of a simultaneous fit on the dimuon mass spectrum, we are finally left with the superposition of the pseudo-proper decay length distributions from prompt and displaced $J/\psi$. The feasibility of the measurement depends therefore on the detector performance in distinguishing the two contributions: preliminary investigations suggest that the ratio between displaced and prompt production can be measured with an error smaller than $10\,\%$ for $p_\mathrm{T} > 1$~GeV/$c$, while uncertainties as large as $20\,\%$ are expected for $p_\mathrm{T} < 1$~GeV/$c$, due to the smaller S/B, the less favorable displaced/prompt ratio, and the less significantly different decay length distributions.

\section{Open Heavy Flavors}

Heavy flavor production is an important probe of the hot and dense medium produced in ultra-relativistic heavy-ion collisions. Due to their large masses, charm and beauty quarks are produced in the initial hard scattering, which makes them a precious source of information on the early stages of the collision; on the other hand, thanks to their long lifetimes, heavy-flavored hadrons can experience the further stages of the evolution of the medium. 

Measurement of open heavy flavor production represents a primary item and a natural calling for the MFT physics, via the analysis of the offset distributions of single muons and dimuons. It should be remarked that the open heavy flavor measurements possible with the current Muon Arm setup do not allow a separation between the open charm and open beauty components; in addition, they rely on the analysis of the $p_\mathrm{T}$ distributions, implying a dependence either on the assumed models or, for the analysis of the Pb--Pb data, on the extrapolation of the available measurements to the rapidity range of the Muon Arm: in this latter case, no trustworthy result can be obtained below $p_\mathrm{T} \approx 4$~GeV/$c$, because of the uncertainties on the background subtraction. In a scenario including the MFT, the measurement is performed on the basis of the analysis of the muons' offset: indeed, offset distributions can be decomposed $-$ with rather limited model assumptions $-$ into the three expected 
components coming from the background, the open charm and the open beauty processes. Preliminary results have shown that this procedure is reliable down to $p_\mathrm{T} \approx 2$~GeV/$c$ in the single muons analysis; below this limit, the similarities between the offset distributions of the open beauty and background components only allow the open charm to be properly measured. Similar investigations are ongoing for the analysis of the muon pairs, for which encouraging results had been found in ideal detector conditions.

\section{Low Mass Dimuons}

The major interest in studying low mass dimuon production is represented by the possibility to study the bulk properties and the space-time evolution of the hot and dense QCD matter formed in ultra-relativistic heavy-ion collisions, revealing microscopic properties such as the relevant degrees of freedom and the hadronic excitation spectrum in the deconfined medium. The main experimental challenge, in pursuing this research, comes from the necessity to subtract with high accuracy the large combinatorial background coming from semi-muonic decays of pions and kaons, in order the signals of interest to be properly identifiable among the rich superposition of sources populating the low mass dimuon spectrum. 

Here, the MFT is expected to be doubly helpful. Firstly, since we deal with prompt signals, the possibility to apply tight cuts on the offset of the muons and dimuons will enhance the S/B by factors as large as~10, without any loss of significance. Secondly, the much improved mass resolution $-$ by factors larger than~3, thanks to the precise measurement of the opening angle of muon pairs $-$ will enable us to precisely identify the 2-body decays of the narrow light resonances, which could be either subtracted to isolate the signals in the underlying continuum, or taken as the object of dedicated investigations.

\section{Conclusions}

The muon physics program of the ALICE experiment is conditioned by the intrinsic limitations of the current setup of the Muon Arm. In this report we discussed how the addition of a silicon Muon Forward Tracker in the acceptance of the Muon Spectrometer should overcome these limitations increasing the physics potential of the muon measurements in ALICE. In particular, the rejection power gained against muon pairs from the combinatorial background will allow us to dramatically improve the uncertainties associated to the measurement of the $\psi'$ charmonium state, while the identification of displaced and prompt $J/\psi$ production will be made possible by means of the measurement of the pseudo-proper decay length distributions. The enhanced pointing accuracy gained by the muon tracks will also give us the possibility to disentangle  open charm and beauty production on the basis of the analysis of the single muon and dimuon offset distributions, starting from $p_\mathrm{T} > 2$~GeV/$c$. Low mass 
dimuons, finally, will enormously profit from the significantly improved rejection power for the combinatorial background, as well as from the much better mass resolution available for the low mass narrow resonances.


\section*{References}

\begin{thebibliography}{9}

  \bibitem{Aamodt:2008zz}
  The ALICE Collaboration, {\emph{The ALICE experiment at the CERN LHC}} 
  {JINST \textbf{3}, S08002 (2008)}

  \bibitem{Alice:2012nf}
  The ALICE Collaboration, {\emph{$J/\psi$ suppression at forward rapidity in Pb--Pb collisions at $\sqrt{s_\mathrm{NN}} =$ 2.76~TeV}}\\ {Phys.~Rev.~Lett.~\textbf{109}, 072301 (2012)}  
  
  \bibitem{Alice:2012fn}
  The ALICE Collaboration, {\emph{$J/\psi$ polarization in pp collisions at $\sqrt{s} = $ 7~TeV}}\\
  {Phys.~Rev.~Lett.~\textbf{108} 082001 (2012)}
  
  \bibitem{Alice:2012aa}
  The ALICE Collaboration, {\emph{Production of muons from heavy flavor decays at forward rapidity in pp and Pb--Pb collisions at $\sqrt {s_{\rm NN}} = $ 2.76~TeV}} {Phys.~Rev.~Lett.~\textbf{109}, 112301 (2012)}
  
  \bibitem{Alice:2012bb}
  The ALICE Collaboration, {\emph{Heavy flavor decay muon production at forward rapidity in proton--proton collisions at $\sqrt{s}$ = 7 TeV}}
  {Phys.~Lett.~B \textbf{708}, 265 (2012)}

  \bibitem{Alice:2012cc}
  The ALICE Collaboration, {\emph{Light vector meson production in pp collisions at $\sqrt{s}$ = 7 TeV}}\\ 
  {Phys.~Lett.~B \textbf{710}, 557 (2012)}
  
\end{thebibliography}
\end{document}